# Fast Integral Histogram Computations on GPU for Real-Time Video Analytics

Mahdieh Poostchi, Kannappan Palaniappan, Da Li , Michela Becchi, Filiz Bunyak and Guna Seetharaman

**Abstract**—In many Multimedia content analytics frameworks feature likelihood maps represented as histograms play a critical role in the overall algorithm. Integral histograms provide an efficient computational framework for extracting multi-scale histogram-based regional descriptors in constant time which are considered as the principle building blocks of many video content analytics frameworks. We evaluate four different mappings of the integral histogram computation onto Graphics Processing Units (GPUs) using different kernel optimization strategies. Our kernels perform cumulative sums on row and column histograms in a cross-weave or wavefront scan order, use different data organization and scheduling methods that is shown to critically affect utilization of GPU resources (cores and shared memory). Tiling the 3-D array into smaller regular data blocks significantly speeds up the efficiency of the computation compared to a strip-based organization. The tiled integral histogram using a diagonal wavefront scan has the best performance of about 300.4 frames/sec for $640 \times 480$ images and 32 bins with a speedup factor of about 120 using GTX Titan X graphics card compared to a single threaded sequential CPU implementation. Double-buffering has been exploited to overlap computation and communication across sequence of images. Mapping integral histogram bins computations onto multiple GPUs enables us to process 32 giga bytes integral histogram data (of 64MB Image and 128 bins) with a frame rate of 0.73 Hz and speedup factor of 153X over single-threaded CPU implementation and the speedup of 45X over 16-threaded CPU implementation.

**Index Terms**—multimedia content analysis, real-time processing, integral histogram, dual-buffering, GPUs.

## 1 INTRODUCTION

Multimedia content analysis are becoming more popular and complex with the advent of accessible high quality camera sensors and various video analytics applications that enable us to automatically process a video to characterize and extract temporal and spatial information for different functionalities. In many video analytics frameworks feature likelihood maps represented as histograms play a critical role in the overall algorithm. Histogram-based regional descriptors are fundamental building blocks of many multimedia analysis tasks including filtering [1], [2], [3], classification and recognition [4], [5], video content retrieval [6], [7], [8], object detection [9], [10], visual tracking [11], [12], [13], 3D reconstruction [14], [15] and video surveillance systems [16], [17].

A histogram is a discretized probability distribution where each bin represents the probability of observing a specific range of feature values for a given descriptor such as intensity, color, edginess, texture, shape or motion. Robustness to geometric deformations makes histogram-based feature representation appealing for many applications. In [18], Porikli generalized the concept of integral image and presented computationally very fast method to extract the histogram of any arbitrary region in constant time. Integral histogram provides an optimum and complete solution for the multi-scale histogram-based search problem. The sequential implementation of the integral histogram uses an $O(N)$ recursive row-dependent method, for an image with $N$ pixels. Once the integral histogram has been computed then a histogram over any arbitrary contiguous rectangular region can be computed in $O(1)$ constant time [18], [19].

We presented two approaches to accelerate the integral histogram computations using GPUs [20] to meet the demands of real-time video processing particularly for large scale and high throughput image sequences: the naive and the efficient implementations which are based on the crossweave scanning method and are utilizing the available Nvidia CUDA SDKs [21], [22]. Mapping video content analysis tasks onto many core GPUs has benefits ranging from faster processing, deeper search, greater scalability, and better performance for multimedia content visualization and encoding [23], [24], classification and recognition [25], [26], [27], object detection and visual tracking [28], [29], etc..

[20] presents how a naive approach with multiple scan suffers from too many kernel invocations, high communication overhead and low GPU utilization. Therefore, the efficient single scan is proposed to reduce the communication overhead by copying the image to GPU with a single transaction and generating the integral histogram over GPUs; however no effort has been done to optimize the integral histogram kernel computations.

This paper pursues exploring different techniques to efficiently compute integral histograms on GPU architecture using the NVIDIA CUDA programming model [32], [33]. The advent of CUDA and its related GPU-accelerated libraries (Thrust [34], cuBLAS [35], MAGMA [36] or cuDNN [37]) help programmers to rapidly develop complex applications using highly-optimized algorithms and functions.

• M. Poostchi, K. Palaniappan, D. Li, M. Becchi and F. Bunyak are with the Department of Electrical Engineering and Computer Science, University of Missouri-Columbia, Columbia, MO, 65211. e-mail: mpoostchi@mail.missouri.edu, palaniappank@missouri.edu, dlx7f@mail.missouri.edu, becchim@missouri.edu, bunyak@missouri.edu
• G. Seetharaman is with Air Force Research Laboratory, Rome, NY 13441.

CUDA kernels outperform the OpenCL portable codes and has more comprehensive documentations and sources [38], [39], [40].

The contributions of this work can be summarized as follows:

- We described four GPU implementations of the integral histogram: Cross-Weave Baseline (CW-B), Cross-Weave Scan-Transpose-Scan (CW-STS), customized Cross-Weave Tiled horizontal-vertical Scan (CW-TiS) and Wave-Front Tiled Scan (WF-TiS). All the implementations rely on parallel cumulative sums on row and column histograms. The first three designs operate a cross-weave scan, and the latter does a wave-front scan.

- In our implementations, we show a trade-off between productivity and efficiency. In particular, the less efficient CW-B and CW-STS solutions rely on existing open-source kernels, whereas the most efficient CW-TiS and WF-TiS designs are based on custom parallel kernels. We show analogies between the computational pattern of the integral histogram and that of the bioinformatics Needleman-Wunsch algorithm. We leverage their similarity to design our most efficient implementation (WF-TiS).

- We relate the performances of our proposed implementations to their utilization of the underlying GPU hardware, and use this analysis to gradually improve over the naive CW-B scheme.

- When computing the 32-bin integral histogram of a $512 \times 512$ image, our custom implementation WF-TiS achieves a frame rate of 135 $fr/sec$ on Tesla K40c (Fig. 15(c)) and 351 on a GeForce GTX Titan X graphics card (Fig. 15(d)). Further, our GPU WF-TiS design reports a 60X speedup over a serial CPU implementation, and a 8X to 30X speedup over a multithreaded implementation deployed on an 8-core CPU server (Fig. 19(b)).

- We have exploited task-parallelism to overlap computation and communication across the sequence of images. Using dual-buffering improved the performance by a factor of two when computing 16-bins integral histogram of HD ($1280 \times 720$) images versus no dual-buffering on GeForce GTX 480 (Fig. 13).

- We evaluated utilizing multiple GPUs for large scale images due to the limited GPU global memory. The integral histogram computations of different bins are distributed across available GPUs using a task queue. We achieved the increasing speedup range from 3X for HD images to 153X for the large 64MB images and 128 bins over the single threaded CPU implementation (Fig. 17).

The remainder of this paper is organized as follows: Section 2 provides background on the integral histogram and on the GPU architecture. Section 3 describes the integral histogram data structure, its layout in GPU memory, and different implementation strategies. A performance evaluation of the proposed kernel implementations, exploiting dual-buffering and utilizing multiple GPUs are presented in section 4.

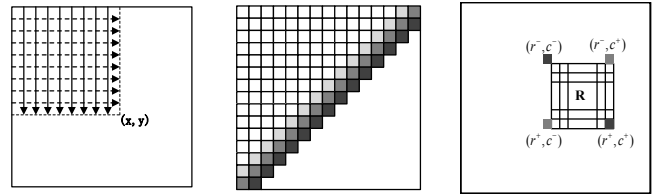

Fig. 1: Computation of the histogram up to location $(x, y)$ using a cross-weave horizontal and vertical scan on the image (left) or wave-front scan (middle). Computation of the histogram for an arbitrary rectangular region $R$ (origin is the upper-left corner with y-axis horizontal(right)).

## 2 BACKGROUND

In this section, we provide a description of the integral histogram computation for 2D arrays as well as a brief summary of the key architectural features of the current generation of NVIDIA GPUs.

### 2.1 Integral Histogram

The integral histogram is a recursive propagation preprocessing method used to compute histograms over arbitrary rectangular regions in constant time [18]. The efficiency of the integral histogram approach enables real-time histogram-based exhaustive search in vision tasks such as object recognition. The integral histogram is extensible to higher dimensions and different bin structures. The integral histogram at position $(x, y)$ in the image holds the histogram for all the pixels contained in the rectangular region defined by the top-left corner of the image and the point $(x, y)$ as shown in Figure 1. The integral histogram for a rectangular 2D region defined by the spatial coordinate $(x, y)$ and bin variable $b$ is defined as:

$$H(x, y, b) = \int_0^x \int_0^y Q(I(r, c), b) dr\, dc \approx \sum_{r=0}^x \sum_{c=0}^y Q(I(r, c), b) \tag{1}$$

where $Q(I(r, c), b)$ is the binning function that evaluates to 1 if $I(r, c) \in b$ for the bin $b$, and evaluates to 0 otherwise. Sequential computation of integral histograms $H$ is described in Algorithm 1. Given $H$, computation of the histogram for any test region $R$ (Fig. 1) reduces to the combination of four integral histograms:

$$\begin{aligned} h(R, b) = \ &H(r^+, c^+, b) - H(r^-, c^+, b) \\ &- H(r^+, c^-, b) + H(r^-, c^-, b) \end{aligned} \tag{2}$$

Figure 1 illustrates the notation and accumulation of integral histograms using vertical and horizontal cumulative sums (prescan), which is used to compute regional histograms.

### 2.2 GPU Architecture

The Graphics Processing Units, invented by Nvidia in 1999, have outperform CPUs in arithmetic throughput and memory bandwidth and became an ideal processors to accelerate many data or compute intensive parallel applications. Many efforts have been made to exploit GPUs for non-graphical applications since 2003 and result in developing several graphics APIs for general purpose computing (GPGPUs). In order to facilitate application development and alleviate

**Algorithm 1:** Sequential Integral Histogram

---

**Input** : Image $\mathbf{I}$ of size $h \times w$, number of bins b
**Output** : Integral histogram tensor $\mathbf{H}$ of size $b \times h \times w$
1: **Initial H:**
    $\mathbf{H} \leftarrow 0$
2: **for** k=1:b **do**
3:    **for** x=1:h **do**
4:       **for** y=1:w **do**
5:          $H(k, x, y) \leftarrow H(k, x - 1, y) + H(k, x, y - 1)$
            $- H(k, x - 1, y - 1) + Q(k, I(x, y))$
6:       **end for**
7:    **end for**
8: **end for**

---

the need of having deep understanding of GPU architecture and memory hierarchy, Nvidia introduces the unified graphics and compute architecture (from initial version of *G80* to the latest PASCAL architecture) and the high level programming language *CUDA* [41].

With the launch of *Fermi* architecture in 2009, Nvidia provided the first hybrid computing model where CPUs and GPUs work together to balance the computationally-intensive workloads. The main advantages of *Fermi* over the older G80 and GT200 are Data locality, Double precision performance, ECC memory support, Up to 16 concurrent kernels and C++ support [41]. Later in 2012, Nvidia released *Kepler* architecture equipped with SMXs that each consists of 192 cores and delivers three times more processing performance when compared to a 32-cores *Fermi* streaming multiprocessors (SMs). *Kepler* architecture increases GPU utilization and simplifies parallel program design through its new features including dynamic parallelism, Hyper-Q, flexible grid management and Nvidia GPUDirect$^{TM}$ [42].

Later in 2014, Nvidia's unveiled the advanced CUDA GPU architecture, named as *Maxwell*. This new architecture exceeds *Kepler*'s efficient SMX configuration through improvements in control logic partitioning, workload balancing, clock-gating granularity, compiler-based scheduling and the number of instructions issued per clock cycle. It provides larger dedicated shared memory up to 64KB that is unlike *Fermi* and *Kepler* which partitioned 64KB fast memory between L1 cache and shared memory [43].

Recently, Nvidia released *Pascal* architecture that is designed for the best scalability and is approximately 10X faster than *Maxwell*. *Pascal* GPUs are equipped with HBM2 memory and NVLink Bandwidth to improve unified virtual addressing scheme across the GPUs and between CPUs and GPUs. These features allow data to be transferred 5% to 12% faster across CPUs and GPUs compared to PCI-Express [44].

We summarize the important aspects of the GPU massive parallelism, programmability and memory hierarchy that is pertinent to the efficient use of GPU kernels and structures of the integral histogram computation flow for large images.

### 2.2.1 Massive parallelism and programmability

NVIDIA GPUs consist of several Streaming Multiprocessors, each containing a set of in-order cores. A *Fermi* SM comprises either 32 or 48 cores; a *Kepler* SMX consists of 192 single-precision CUDA cores; and the maxwell architecture has 128 CUDA cores per SMM.

The advent of the Compute Unified Device Architecture (CUDA) has greatly improved the programmability of NVIDIA GPUs. With CUDA, the computation is organized in a hierarchical fashion, wherein threads are grouped into thread blocks. Each thread block is mapped onto a SM, whereas different threads are mapped to cores and executed in SIMD units. Threads within the same block can communicate using shared memory, whereas threads within different thread blocks are fully independent. Therefore, CUDA exposes to the programmer two degrees of parallelism: fine-grained parallelism within a thread block and coarse-grained parallelism across multiple thread blocks. GPU utilization is maximized when threads belonging to the same warp do not present divergent control flows, and when the kernel launch configuration (number of threads and thread blocks) is such to fully utilize the underlying cores.

### 2.2.2 GPU memory hierarchy

GPUs have a heterogeneous memory organization consisting of high latency off-chip global memory, low latency read-only constant memory (which resides off-chip but is cached), low-latency on-chip read-write shared memory, and texture memory. GPUs adopting the *Fermi*, *Kepler* and *Maxwell* architecture, such as those used in this work, are also equipped with a two-level cache hierarchy. Judicious use of the memory hierarchy and of the available memory bandwidth is essential to achieve good performances.

The global memory can be accessed via 32-, 64- or 128-bytes transactions. Multiple memory accesses to contiguous memory locations can be automatically coalesced into a single memory transaction: memory coalescing is fundamental to optimize the memory bandwidth utilization. Further, context-switch among threads can be used to hide high latency global memory accesses. Finally, the use of shared memory and caches can be used to reduce the accesses to global memory.

The shared memory can be configured either as a software- or as a hardware-managed cache. The first configuration is typically for highly optimized code with customized cache management. In this case, within kernel functions, threads will first load the data from global memory to shared memory, then process the data in shared memory, and finally move the results back into global memory. Since shared memory does not impose the same coalescing rules as global memory, it can allow efficient irregular access patterns. In order to achieve high bandwidth, shared memory is divided into equally sized banks. Memory requests directed to different banks can be served in parallel, whereas requests to the same bank are serialized. Therefore, avoiding bank conflicts is essential to optimize shared memory access. *Kepler* architecture address the problem by doubling the bank width of shared memory compared to *Fermi* which was not that much efficient. Unlike *Fermi* and *Kepler* architectures, in which 64KB memory is partitioned between L1 cache and shared memory, *Maxwell* provides up to 64 KB dedicated shared memory to achieve boost the performance [45] and finally *Pascal* innovative memory design deploys Chip-on-Wafer-on-Substrate (CoWoS) with HBM2 to support Big Data Workloads [44].

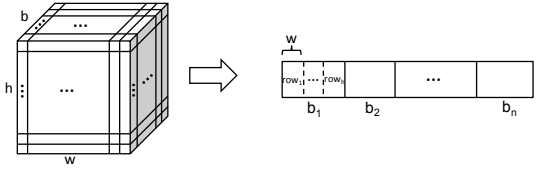

Fig. 2: Integral histogram tensor represented as 3-D array data structure (left), and equivalent 1-D array mapping (right).

# 3 KERNEL OPTIMIZATION FOR INTEGRAL HISTOGRAM

In this section, we first describe the integral histogram data structure and its layout in GPU memory and then present different optimization strategies. In our first implementations [20], we reuse existing parallel kernels from the NVIDIA Software Development Kit (SDK); we refer to these as generic kernels. We point out the limitations of such an approach, and progressively refine our implementation in order to better utilize the architectural features of the GPU. This leads to the evolution of four techniques to compute the integral histogram on GPUs that trade-off productivity with efficiency and a discussion of how the performance of the proposed implementations reflect their utilization of the underlying hardware. The first three implementations perform cumulative sums on row and column histograms in a cross-weave (CW) fashion, whereas the fourth one performs a wavefront (WF) scan.

## 3.1 Data Structure Design

An image with dimensions $h \times w$ produces an integral histogram tensor of dimensions $b \times h \times w$, where $b$ is the number of bins in the histogram. This tensor can be represented as a 3-D array, which in turn can be mapped onto a 1-D row major ordered array as shown in Figure 2. It is well known that the PCI-express connecting CPU and GPU is best utilized by performing a single large data transfer rather than many small data transfers. Therefore, whenever the 1-D array representing the integral histogram fits the available GPU global memory, we transfer it between GPU and CPU with a single memory transaction. The computation of larger integral histograms is tiled along the bin-direction and distributed between available GPUs: portions of the 1-D array corresponding to the maximum number of bins that fit the GPU capacity are transferred between GPU and CPU in a single transaction. For all the considered image sizes a single bin fits the GPU memory; however, our implementation can be easily extended to images exceeding the GPU capacity by tiling the computation also column-wise. Finally, we experimentally verified that initializing the integral histogram on GPU is more efficient than initializing it on CPU and then transferring it from CPU to GPU. Therefore, in all our GPU implementations, we initially transfer the image from CPU to GPU, then initialize and compute the integral histogram on GPU, and finally transfer it back from device to host.

## 3.2 CW-B: Naive Cross-weave Baseline Parallelization (Using CUDA SDK)

The sequential implementation of the integral histogram is represented by Algorithm 1. As can be seen, the algorithm can be trivially parallelized along the $b$-dimension, since the computation of different bins can be done fully independently. However, the algorithm presents loop-carried dependences along both the $x-$ and the $y-$ dimensions. Therefore, we need an intelligent mechanism for inter-row and inter-column parallelization. The cross-weave scan mode (Fig. 1) enables cumulative sum tasks over rows (or columns) to be scheduled and executed independently allowing for inter-row and column parallelization. We observe that NVIDIA CUDA SDK provides an efficient implementation of the *all-prefix-sums* [21] and of the *2-D transpose* [22] operations. Therefore, leveraging these existing open source kernels, we can quickly implement the integral histogram. Algorithm 2 presents our cross-weave baseline approach (CW-B) to compute the integral histogram, combining cross-weave scan mode with the existing parallel prefix sum and 2-D transpose implementations. In this approach, we first apply prefix-sums to the rows of the histogram bins (horizontal cumulative sums or prescan), then transpose the array corresponding to each bin using a 2-D transpose, and finally reapply the prescan to the rows of the transposed histogram to obtain the integral histograms at each pixel. We now briefly describe the two parallel kernels available in NVIDIA SDK.

### 3.2.1 Parallel Prefix Sum Operation on the GPU

The core of our CW-B approach is the parallel prefix sum algorithm [21]. The *all-prefix-sums* operation (also refered as *scan*) applied to an array generates a new array where each element $k$ is the sum of all values preceding $k$ in the scan order. Given an array $[a_0, a_1, ..., a_{n-1}]$ the prefix-sum operation returns

$$[0, a_0, (a_0 + a_1), ..., (a_0 + a_1 + ... + a_{n-2})] \tag{3}$$

The parallel prefix sum operation on GPU consists of two phases: an *up-sweep* and a *down-sweep* phase (see Fig. 3). The *up-sweep* phase builds a balanced binary tree on the input data and performs one addition per node. Scanning is done from the leaves to the root. In the *down-sweep* phase the tree is traversed from root to the leaves and partial

---

**Algorithm 2:** CW-B: Naive Cross-weave Baseline Parallelization

**Input :** Image **I** of size $h \times w$, number of bins b
**Output :** Integral histogram tensor **IH** of size $b \times h \times w$

1: **Initialize IH**
   $\mathbf{IH} \leftarrow 0$
   $\mathbf{IH}(\mathbf{I}(\mathbf{w}, \mathbf{h}), \mathbf{w}, \mathbf{h}) \leftarrow 1$
2: **for** all bins b **do**
3:     **for** all rows x **do**
4:         //horizontal cumulative sums
       $IH(x, y, b) \leftarrow IH(x, y, b) + IH(x, y - 1, b)$
5:     **end for**
6: **end for**
7: **for** all bins b **do**
8:     //transpose the bin-specific integral histogram
       $IH^T(b) \leftarrow$ 2-D Transpose($IH(b)$)
9: **end for**
10: **for** all bins b **do**
11:     **for** all rows y of $IH^T$ **do**
12:         //vertical cumulative sums
        $IH^T(y, x, b) \leftarrow IH^T(y, x, b) + IH^T(y, x - 1, b)$
13:     **end for**
14: **end for**

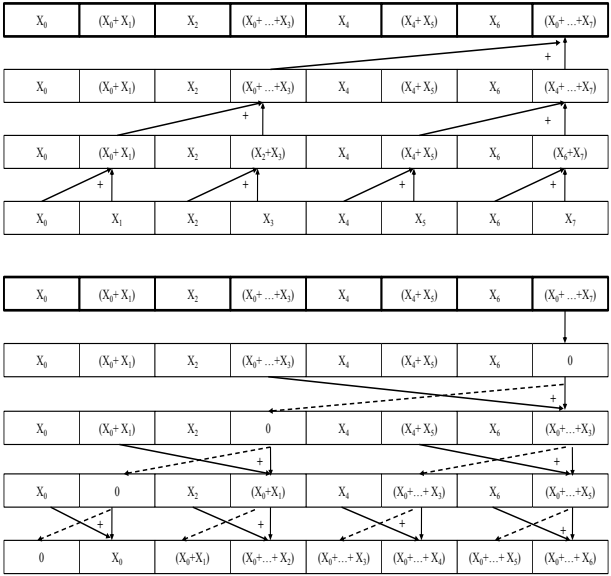

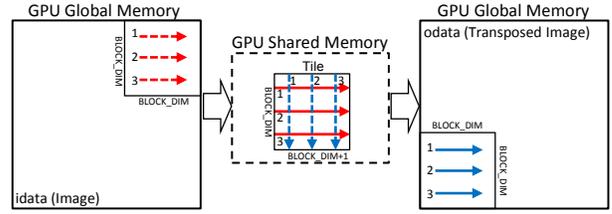

Fig. 4: Data flow between GPU global memory and shared memory within the transpose kernel; stage 1 in red, stage 2 blue, reads are dashed line, writes are solid lines.

Fig. 3: Parallel prefix sum operation, commonly known as exclusive scan or prescan [21]. Top: Up-sweep or reduce phase applied to an 8-element array. Bottom: Down sweep phase.

sums from the up-sweep phase are aggregated to obtain the final scanned (prefix summed) array. Prescan requires only $O(n)$ operations: $2 \times (n-1)$ additions and $(n-1)$ swaps. Padding is applied to shared memory addresses to avoid bank conflicts.

### 3.2.2 GPU-based 2D Transpose Kernel

The integral histogram computation requires two prescans over the data: a horizontal prescan that computes cumulative sums over rows of the data, followed by a vertical prescan that computes cumulative sums over the columns of the first scan output. Taking the transpose of the horizontally prescanned image histogram enables us to reapply the same (horizontal) prescan algorithm effectively on the columns of the data. The transpose operation can be performed using the efficient 2D kernel described in [22]. This tiled implementation uses shared memory to avoid uncoalesced read and write accesses to global memory, and uses padding to avoid shared memory conflicts and thereby optimize the shared memory accesses.

### 3.3 CW-STS: Cross-weave Scan-Transpose-Scan Parallelization (Using CUDA SDK)

As can be observed in Algorithm 2, the CW-B implementation performs many kernel invocations: $b \times h$ times horizontal scans each of size $w$, $b$ times 2-D transposes each of size $w \times h$, and $b \times w$ times vertical scans each of size $h$. In other words, this approach is based on the use of many parallel kernels, each of them performing very little work and therefore greatly under-utilizing the GPU. With the regular image sizes in consideration, for example, the scan kernels are invoked on arrays of size varying from 512 to 2048; however, the all-prefix-sum kernel has been designed to perform well on arrays consisting of millions of elements. Therefore, an obvious way to improve the integral histogram implementation is to increase the amount of work performed by each kernel invocation, and reduce kernel invocation overheads. This, in turn, will improve the GPU utilization. Luckily, as shown in Algorithm 3, the computation can be easily broken into three phases: a single horizontal scan, a single 3-D transpose, and a single vertical scan. We call this solution cross-weave scan-transpose-scan (CW-STS) parallelization. We observe that the CW-STS approach does not require rewriting the prescan kernel, but only invoking it judiciously. As explained above, the all-prefix-sum operation consists of two phases: the up-sweep computes partial sums, and the down-sweep aggregates them into the final result. However, the problem is that in the early stages of the up-sweep phase much more threads are involved in the computation than in the later stages (leading to GPU under-utilization). The down-sweep phase has the same problem except that the thread utilization, initially minimal, increases with the computational stage (Figure 3). To avoid this problem, we proposed the CW-TiS and WF-TiS approaches which compute all row- or column-wise scans with a single kernel call. Specifically, in the horizontal and vertical scan phase we invoke the custom prescan kernel *once* using a 2-D grid of size $(b, \frac{w \times h}{2 \times Num\_Threads})$.

In order to allow a single transpose operation, we need to transform the existing 2-D transpose kernel into a 3-D transpose kernel. This can be easily done by using the bin offset in the indexing. The 3-D transpose kernel is launched using a 3-D grid of dimension $(b, \frac{w}{BLOCK\_DIM}, \frac{h}{BLOCK\_DIM})$, where BLOCK_DIM is the maximum number of banks in shared memory (32 for all graphics card used). Figure 4 shows the data flow in the transpose kernel. A tile of size BLOCK_DIM * BLOCK_DIM is written to the GPU shared memory into an array of size BLOCK_DIM*(BLOCK_DIM+ 1). This pads each row of the 2-D block in shared memory so that bank conflicts do not occur when threads address the array column-wise. Each transposed tile is written back to the GPU global memory to construct the full histogram

---

**Algorithm 3:** CW-STS: Single Scan-Transpose-Scan Parallelization

**Input** : Image **I** of size $h \times w$, number of bins b
**Output** : Integral histogram tensor **IH** of size $b \times h \times w$
1: **Initialize IH**
    **IH** $\leftarrow 0$
    **IH**$(\mathbf{I}(\mathbf{w}, \mathbf{h}), \mathbf{w}, \mathbf{h}) \leftarrow 1$
2: **for** all $b \times h$ blocks in parallel **do**
3:     Prescan($IH$)
4: **end for**
5:   //transpose the histogram tensor
    $IH^T \leftarrow$ 3D Transpose($IH$)
6: **for** all $b \times w$ blocks in parallel **do**
7:     Prescan($IH^T$)
8: **end for**

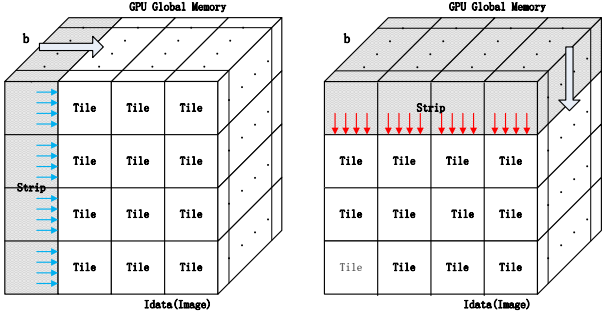

Fig. 5: Cross-weave Tiled Horizonal-Vertical Scan (CW-TiS): Tiled horizontal scan (left), Tiled vertical scan (right). Both scans are performed in tile and the shadow area presents one strip.

transpose.

### 3.4 CW-TiS: Cross-weave Tiled Horizonal-Vertical Scan Parallelization

To understand how to further improve the integral histogram computation, we can observe the following. First, the use of the transpose kernel in the CW-STS implementation is motivated by the reuse of the prescan kernel in the vertical scan phase. However, the transpose operation can take considerable time compared to the prescan. For instance, when the image size is $512 \times 512$ and the histogram consists of 32 bins, the transpose takes about 20% of the whole computation time, and almost 50% of the time of a single prescan (Figure 8). Therefore, the execution time can be greatly reduced by combining the transpose and the vertical prescan into a single parallel kernel.

Second, the prescan kernel has its own limitations. While in the early stages of the up-sweep phase many threads are involved in the computation, in later stages only a few threads are active. For instance, as can be observed in Fig. 3, the number of active threads decreases from 4 at the beginning to 1 at the end. The down-sweep phase has the same problem (except that the thread utilization, initially minimal, increases with the computational stage). In general, the up-sweep and down-sweep phases of a scan on an array of $n$-element will consist of $2 \times log_2(n)$ iterations. In the up-sweep the number of active threads, initially equal to $\frac{n}{2}$, halves at every iteration. The number of working cycles of all active threads is therefore equal to $3 \times (n-1)$. The efficiency, defined as the ratio of the total working cycles over the product of the number of threads and the number of iterations will be

$$\frac{3 \times (n-1)}{n \log_2 n} \approx \frac{3}{\log_2 n} \quad (4)$$

For example, the efficiency of the scan on a 1024-element 1-D array is only 30%. To achieve a better efficiency, we can leverage the data-level parallelism underlying the integral histogram computation. As mentioned before, different bins can be processed fully independently. In addition, for each bin, the horizontal scan can be performed fully independently on the $h$ rows, and the vertical scan can be performed in parallel on the $w$ columns. Based on these observations, we propose a cross-weave tiled horizontal and vertical scan (CW-TiS) method that: (i) eliminates the need for the transpose operation and benefits less memory, and

(ii) better utilizes the data-level parallelism of the integral histogram. Algorithm 4 represents how CW-TiS operates. First, each of the $b$ matrices of size $(h \times w)$ corresponding to different bins is divided into *tiles*. Each tile must be small enough to fit in shared memory and large enough to contain sufficient amount of data for computation work. In our implementation, we use squared tiles. The processing is divided into two stages: the horizontal scan (Fig. 5 (left)) and the vertical scan (Fig. 5 (right)). In each stage, the computation is performed strip-wise until the whole matrix has been processed. A kernel call operates on one strip of size $tile\_width \times image\_height$ in horizontal scan versus $image\_width \times tile\_height$ during the vertical scan. The number of vertical strips in horizontal scan equals to $VStrips = \frac{w_{Image}}{w_{Tile}}$. Whereas the number of horizontal strips during the vertical scan is $HStrips = \frac{h_{Image}}{h_{Tile}}$. Therefore, The total number of image tiles or blocks being processed is given by:

$$Tiles = \frac{w_{Image} \times h_{Image}}{w_{Tile} \times h_{Tile}} = VStrips \times HStrips \quad (5)$$

We expect the image sizes to be evenly divisible by the tile sizes otherwise the image will be appropriately padded. In the kernel implementation, each thread-block is assigned to a tile and each thread to a row/column. Shared memory is used to allow efficient and coalesced memory accesses. Threads belonging to the same block push the cross-weave forward (either from left to right or from top to bottom (Fig. 5). Since each thread-block consists of warps, in order to avoid thread divergence within warps and GPU underutilization, the tile size is set to be a multiple of the warp size (32).

### 3.5 WF-TiS: Wave-front Tiled Scan Parallelization

The use of separate horizontal and vertical scan kernels in the CW-TiS method has a drawback: it causes each tile to be transferred multiple times between global and shared memory. In fact, in both scan kernels, each tile is first moved from global into shared memory, processed and then moved back to global memory. As a consequence, combining the horizontal and vertical scans into a single kernel will allow accessing global memory only twice per tile (once in read, and once in write mode).

Before introducing the detailed implementation, let us briefly analyze the data dependences of the integral histogram. For the horizontal scan, the data in each row rely on the data on their left; for the vertical scan, the data

---

**Algorithm 4:** CW-TiS: Cross-weave Tiled Horizonal-Vertical Scan Parallelization

**Input** : Image $\mathbf{I}$ of size $h \times w$, number of bins b
**Output** : Integral histogram tensor $\mathbf{IH}$ of size $b \times h \times w$

1: **Initialize IH**
    $\mathbf{IH} \leftarrow 0$
    $\mathbf{IH}(\mathbf{I}(\mathbf{w}, \mathbf{h}), \mathbf{w}, \mathbf{h}) \leftarrow 1$
2: **for** all bins b in parallel **do**
3:   **for** all vertical strips $v_s$ of width $TILE\_SIZE$ **do**
4:     **Tiled_Horizontal_Scan(IH)**
5:   **end for**
6:   **for** all horizontal strips $h_s$ of height $TILE\_SIZE$ **do**
7:     **Tiled_Vertical_Scan(IH)**
8:   **end for**
9: **end for**

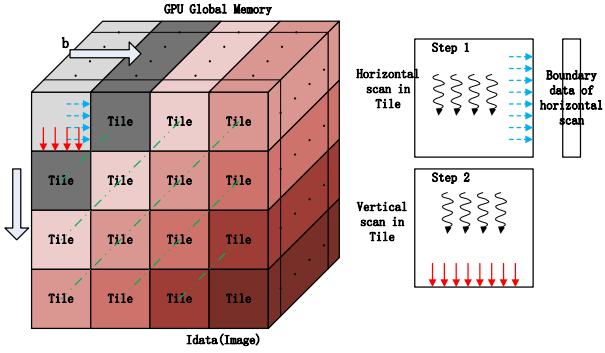

Fig. 6: "WF-TiS" implementation. (Left) Tiles with the same color belong to the same stride and are executed in the same kernel launch iteration. (Right) The horizontal and vertical scan of a tile. It starts from the horizontal scan, store the boundary data into extra memory and finally the vertical scan.

in each column rely on the data on their upper position. This data access pattern is quite similar to that in the GPU implementation of the Needleman-Wunsch algorithm [46] in the Rodinia Benchmark Suite [47]. Therefore, we can arrange the computation in a similar fashion, and compute the integral histogram using a wave-front scan. Algorithm 5 represents the latest approach which we called Wave-front Tiled Scan Parallelization (WF-TiS). Similarly to the CW-TiS implementation, we divide the $h \times w$ matrix into different tiles as shown in Fig. 6. Again, each tile should be small enough to fit in shared memory, and large enough to contain non-trivial amount of computation work. All the tiles lying on the same diagonal line with different bins (they are presented with the same color) are considered part of the same strip and processed in parallel. Therefore, for an image of size $w \times h$ and tile size $TILE\_SIZE$ the total number of iterations are

$$\lceil \frac{w}{TILE\_SIZE} \rceil + \lceil \frac{h}{TILE\_SIZE} \rceil - 1 \qquad (6)$$

Within the parallel kernel, each thread block will process a tile, and each thread will process a row (during horizontal scan) and a column (during the vertical scan) of the current tile. The tricky part of this implementation is that after the horizontal scan, the last column of each tile must be preserved for horizontal scan of the next strip before being overwritten during the vertical scan. This can be achieved by storing the extra data in global memory (the additional required memory is an array of $h$ elements). The WF-TiS method can potentially be preferable to the CW-TiS by eliminating unnecessary data movements between shared and global memory.

## 4 EXPERIMENTAL RESULTS

In this section, we present a performance evaluation of our proposed GPU implementations of the integral histogram. Our experiments were conducted on four GPU cards:

- GeForce GTX Titan X graphics card equipped with $24 \times 128$-core SMM, maxwell architecture, $12GB$ of global memory and compute capability 5.2.

- Nvidia Tesla K40c equipped with $15 \times 192$-core SMX, kepler architecture, 11GB of global memory and compute capability 3.5.
- Nvidia Tesla C2070 equipped with $14 \times 32$-core SMs, fermi architecture and has about 5GB of global memory, compute capability 2.0.
- Nvidia GeForce GTX 480 consists of $7 \times 48$-core SMs, fermi architecture with 1GB global memory and compute capability 2.1.

Our discussion is organized as follows: First, we present a comparative evaluation of the four proposed GPU implementations focusing on the processing time. Second, we show how to tune the parameters of our most efficient implementation to achieve better performances (the WF-Tiled solution). Third, we discuss the impact of the data transfers on the overall performances. Fourth, we exploit double-buffering to overlap the computation and communication across sequences of images and calculate the frame rate of our proposed methods. We have extended our experiments utilizing multiple GPUs for large scale images. To conclude, we compare our GPU implementations using different GPU architectures with the multi-threaded CPU implementation. We tune all kernel functions to achieve the best performances.

### 4.1 Kernel Performance Evaluation

Figure 7 reports the cumulative kernel execution time of the four proposed GPU implementations on different image sizes for a 32-bin integral histogram. For readability, the data in the y-axis are reported in logarithmic scale. As it is obvious, due to its extremely poor GPU utilization, the CW-B approach performs extremely poor, and is outperformed by all other approaches by a factor in excess of 30X. CW-TiS outperforms CW-STS by a factor between 2X and 3X depending on the image size. Finally, WF-TiS leads to a further performance improvement up to about 1.5X over CW-TiS (Fig. 7).

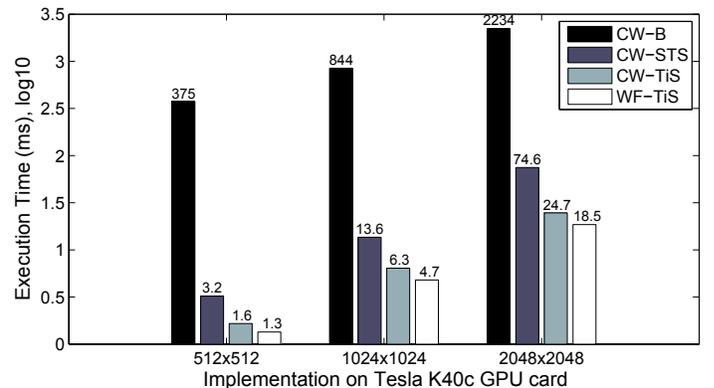

Fig. 7: Cumulative kernel execution time of the four proposed GPU implementations on different image sizes for a 32-bin integral histogram on Tesla K40c.

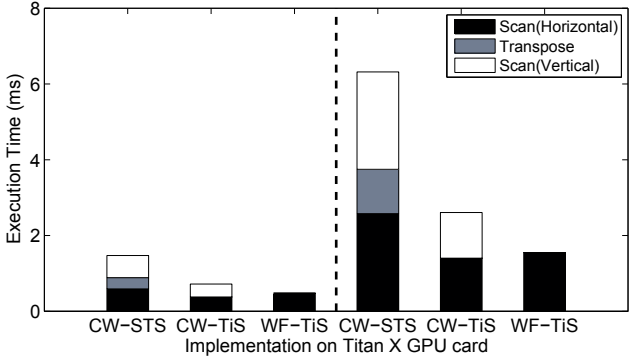

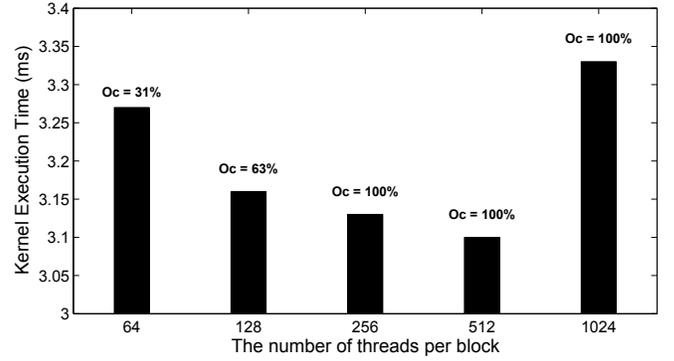

Fig. 8: Kernel execution time for calculating integral histogram of size 512x512x32 (left), and 1024x1024x32 (right) on GTX Titan X.

Fig. 9: Kernel execution time and occupancy for different thread block configurations for an integral histogram of size 512x512x32 on Tesla K40c.

To allow a better understanding of these results, Fig. 8 breaks down the execution times of different processing tasks. We can make the following observations. First, even when performed on large arrays (consisting of $b \times w \times h$ elements), the prescan kernel provided by CUDA SDK is less effective than the custom prescan kernel that we implemented in our CW-TiS and WF-TiS solutions. In fact, the execution time of our custom prescan kernel is comparable to that of the transpose kernel. Second, additional performance boost is achieved by merging the horizontal and the vertical scan into a single scan kernel, reducing the total number of global memory accesses and halving the required memory by removing the transpose phase.

### 4.2 Performance Tuning of WF-TiS

In this section, we discuss the tuning of two important parameters used in our WF-TiS design: the tile size and the thread block configuration. Our considerations can be generalized to other parallel kernels using tiling.

#### 4.2.1 Configuring the Thread Blocks

CUDA kernels can run with different thread block configurations. At runtime, each thread block is mapped onto a SM in a round-robin fashion. The execution of blocks mapped onto the same SM can be interleaved if their cumulative hardware resource requirements do not exceed those available on the SM (in terms of registers and shared memory). In case of interleaved execution, context-switch among thread blocks can help hiding global memory latencies. If interleaved execution is not possible, memory latencies can be hidden only by context switching within a single thread block. Therefore, properly setting the block configuration can help achieving better performances.

NVIDIA provides a 'CUDA Occupancy Calculator' to assist the programmer in finding the kernel configuration that maximizes the resource utilization of the GPU. Although low occupancies (typically below 50%) indicate possible bad performances, a full occupancy does not ensure the optimal configuration. This fact is highlighted in Fig. 9, which shows the kernel execution time and the GPU occupancy using different thread block configurations. These results were reported on a 512×512 image and a 32-bin integral histogram. As can be observed(Fig. 9), both the best and the worst configurations, 512 and 1024 threads, in terms of execution time are characterized by a 100% GPU occupancy. In addition, the lowest execution time is achieved using 512-thread blocks.

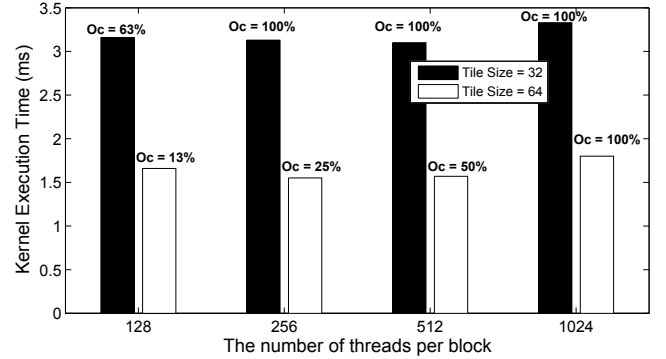

Fig. 10: Performance evaluation of the WF-TS kernel for two tile sizes and several thread block configurations for an integral histogram of size 512x512x32 on Tesla K40c card.

#### 4.2.2 Configuring the Tile Size

The tile size determines the amount of shared memory used by each thread block. Increasing the tile size will reduce the number of iterations (or strips), and will increase the amount of work performed by each thread block. But larger tiles may limit inter-block parallelism and decrease the opportunity to hide global memory latencies by context switching across thread blocks. We tried different tile configurations ($16 \times 16$, $32 \times 32$, $64 \times 64$). However, in the case of $16 \times 16$, the performance is much worse than the two others since the tile data is processed linearly and each line is limited to only 16 elements. This causes that only half of the threads warp be active for each tile. Figure 10 reports a performance analysis of our WF-TiS implementation using two tile sizes ($32 \times 32$ and $64 \times 64$) and several thread block configurations. The $64\times64$ tile configuration performs better than $32 \times 32$ tile configuration by better use of the limited shared memory.

### 4.3 Communication Overhead Analysis

In this section, we discuss the overhead due to data transfers between CPU and GPU. The experiments were performed

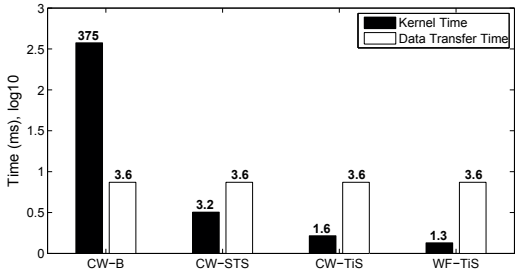

(a) Implementations on *Tesla K40c* 512 × 512, 32 bins

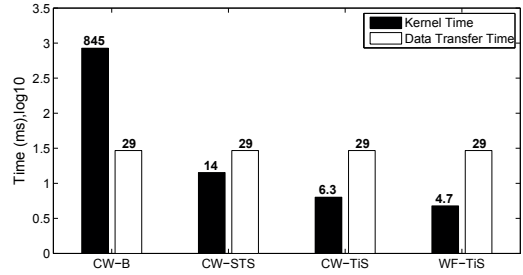

(b) Implementations on *Tesla K40c* 1024 × 1024, 32 bins

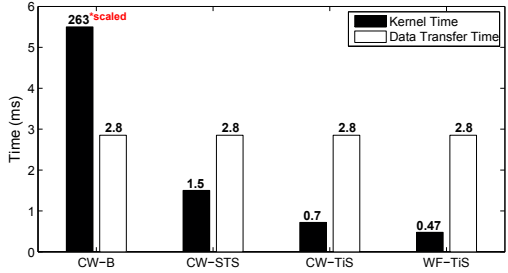

(c) Implementations on *GeForce GTX Titan X* 512 × 512, 32 bins

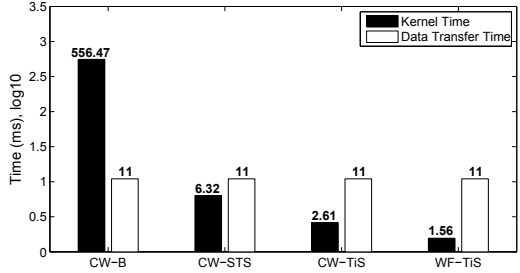

(d) Implementations on *GeForce GTX Titan X* 1024 × 1024, 32 bins

Fig. 11: Kernel execution time versus data transfer time for 64 × 64 tile configuration.

on *Tesla K40c* with kepler architecture and a *GeForce GTX Titan X* with maxwell architecture.

Figure 11 shows the results. We make the following observations for both GPUs: the CW-B implementation is compute bound (that is, the kernel execution time is larger than the CPU-GPU data transfer time), whereas the other solutions are data-transfer-bound. These experiments suggest that further kernel optimizations may not be advisable. On the other hand, potential performance improvements may result from a reduction in the communication overhead coming from advances in interconnection technologies. However, it must be observed that the data transfer overhead is significant only when considering the integral histogram a stand-alone application. *In most cases, the integral histogram is part of a more complex image processing pipeline that can be implemented on GPU*. In these scenarios, since the integral histogram does not need to be transferred back to CPU, optimizing the kernel processing is still relevant.

## 4.4 Overlap Communication and Computation Using Dual-Buffering for Sequence of Images

So far we have focused on distributing integral histogram data across available GPU cores. We can also leverage dual-buffering to overlap CPU-GPU communication and GPU computation across different images for further performance improvement. This can be accomplished by using CUDA streams along with page-locked memory and asynchronous data transfers between host and device. Algorithm 6 represents the pseudo-code of our dual-buffering based solution. All operations issued to stream 1 are independent of those issued to stream 2. In each iteration, the following operations are performed: first, two images are read from disk and stored in page-locked host memory; second, those images are transferred to the GPU

---

**Algorithm 6:** Integral histogram computation on sequences of images using dual-buffering

```
 1: for all pair of images do
 2:    /* Read two images from disk to CPU */
 3:    h_Img1 = CopyImageFromDisk(Img1);
 4:    h_Img2 = CopyImageFromDisk(Img2);
 5:
 6:    /* Transfer two images from CPU to GPU */
 7:    cudaMemcpyAsync(d_Img1, h_Img1, stream1);
 8:    cudaMemcpyAsync(d_Img2, h_Img2, stream2);
 9:
10:    /* Generate and intialize the histograms on GPU */
11:    init_kernel(d_IntHist1, d_Img1, stream1);
12:    init_kernel(d_IntHist2, d_Img2, stream2);
13:
14:    /* Compute integral histograms */
15:    IntHistComputation_kernel(d_IntHist1, stream1);
16:    IntHistComputation_kernel(d_IntHist2, stream2);
17:
18:    /* Transfer integral histogram from GPU to CPU */
19:    cudaMemcpyAsync(h_IntHist1, d_IntHist1, stream1);
20:    cudaMemcpyAsync(h_IntHist2, d_IntHist2, stream2);
21: end for
```

with asynchronous data transfers; then, the initialization and histogram computation kernels are launched; finally, the computed integral histograms are copied back to CPU (again, using asynchronous data transfers). We enqueue the memcpy and kernel execution operations breadth-first across streams rather than depth-first to avoid blocking the copies or kernel executions of a stream with another stream. This sequence of operations allows effective overlapping between different CUDA streams operations. Figure 12 elaborates pipeline parallelism to address the communication and computation overlap when computing HD images 32-bins integral histogram.

Our experiments show that the memory copies from disk to host are twice as fast as the GPU operations. Therefore,

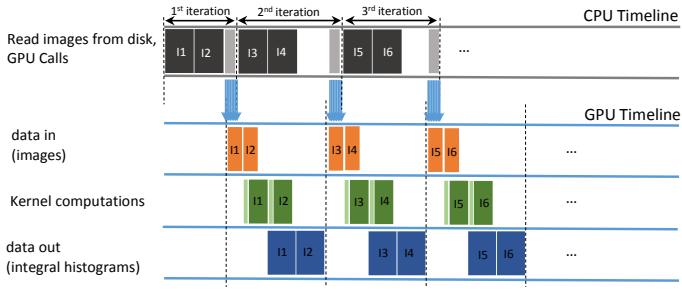

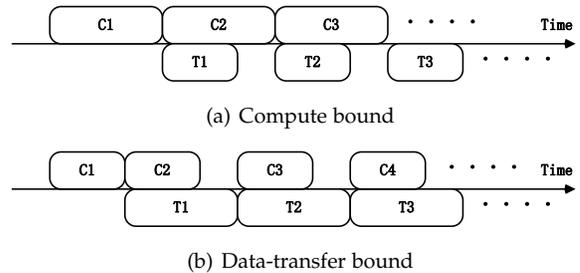

Fig. 12: Pipeline parallelism: communication and computation overlap using dual-buffering.

Fig. 14: Overlapping of computation and communication using double buffering for (a) compute bound and (b) data-transfer bound implementations. In the diagram, $Ci$ and $Ti$ shown in blocks represent kernel computation and data transfer, respectively, for the $i$th integral histogram

in our implementation, such copies are completely hidden behind the GPU tasks. Figure 13 shows the effect of dual-buffering on the frame rate for a sequence of 100 HD images ($1280 \times 720$) using the WF-TiS kernel. As can be seen, dual-buffering improves the performance by a factor of two for 16-bins integral histogram computations. However, as the number of bins increases, the performance improvement decreases, and becomes negligible at 128 bins. This can be attributed to the fact that the use of page-locked memory on very large memory regions leads to performance degradation.

### 4.5 Frame Rate

The frame rate is defined as the maximum number of images processed per second. In this section, we compute the frame rate in the assumption that the integral histogram is a stand-alone application (rather than part of a more complex image processing pipeline on GPU). We consider data transfers between CPU and GPU, but since we use double buffering, the processing of the current image and the transfer of the previously computed integral histogram from GPU to CPU can be overlapped (see Figure 14). Therefore, the frame rate equals to the $(cumulative\_kernel\_execution\_time)^{-1}$ for compute-bound implementations, and to $(data\_transfer\_time)^{-1}$ for data-transfer-bound ones. Fig. 15 $a$ and $b$ show the frame rate for different image sizes on the considered GPUs. The GTX Titan X allows faster data transfers between CPU and GPU. All data are reported on 32-bins integral histograms. In Fig 15 $a$ and $b$, the CW-STS, CW-TiS and WF-TiS implementations are data-transfer-bound for considered images. Figure 15 $c$ and $d$ show the frame rate reported on $512 \times 512$ images with different number of bins. As can be seen in Fig.

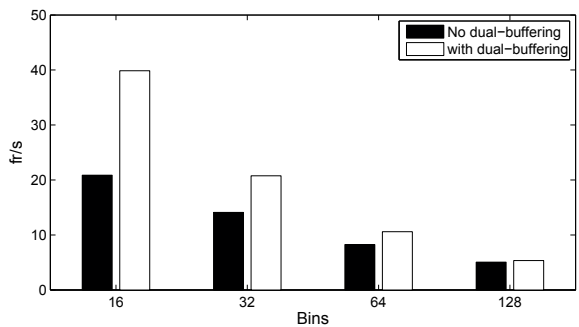

Fig. 13: Effect of dual-buffering on the frame rate of a sequence of 100 HD images ($1280 \times 720$) and different bin sizes. The experiments are conducted using the WF-TiS kernel.

$15c$, the three best implementations are data-transfer-bound. Since increasing the number of bins means increasing the amount of data to be transferred (from GPU to CPU), the performances degrade linearly with the number of bins.

### 4.6 Integral Histogram for Large Scale Images Using Multiple GPUs

We have extended our WF-TiS kernel implementation to process large images (e.g. WHSXGA: $6400 \times 4800$). For such large images, limited GPU global memory becomes the bottle-neck since it can not hold the whole integral histogram with all bins. In our approach, bins are equally grouped into distinct tasks which are enqueued. Each time CPU picks one task from the queue and issues a kernel to the available GPU in the system. Each kernel computes the integral histogram of task bins using the WF-TiS approach. When a device is available, CPU will dispatch another task from the queue to GPU and meanwhile, results will be copied back to CPU. This iterative process will continue until the queue is empty. The experiments are conducted on our superserver equipped with 4 GTX 480 GPUs (Fig. 18). In our system, tasks will be distributed evenly. For instance, if there are 64 bins, each set of 16 bins will be performed on one of the GPUs. The portion of the 16 bins that fit the GPU capacity will be performed in parallel. The most efficient WF-TiS kernel is invoked to compute the integral histogram for each bin.

Figure 16(a) shows the frame rate when computing 32-bins integral histogram for selected high definition image sizes (HD: $1280 \times 720$, FHD: $1920 \times 1080$, HXGA: $4096 \times 3072$, WHSXGA: $6400 \times 4800$ and 64MB: $8k \times 8k$). Figure 16(b) represents the frame rate for HD and FHD images with different number of bins.

Figure 17 shows the speedup of computing 128 bins integral histogram. The increasing speedup range from 3X (for HD) to 153X for the large 64MB images and 128 bins (total 32GB of 4 byte integer) over a single threaded CPU implementation is achieved. There are several advantages of this approach:

- It is an easy-to-scale approach which can utilize multiple GPUs in the node
- The computation time (on GPU) is overlapped with communication time (copy result) via dual-buffering

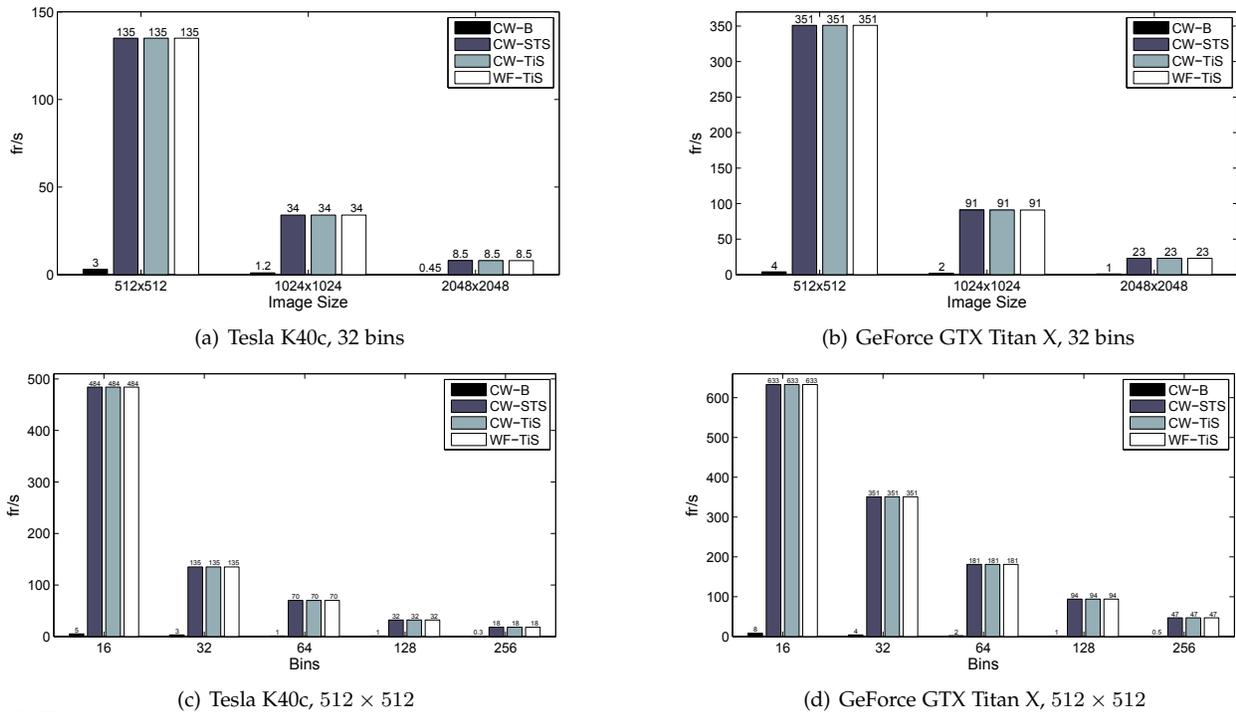

(a) Tesla K40c, 32 bins

(b) GeForce GTX Titan X, 32 bins

(c) Tesla K40c, $512 \times 512$

(d) GeForce GTX Titan X, $512 \times 512$

Fig. 15: Frame rate for 64x64 tile configurations: (a) and (b) for different image sizes, (c) and (d) for $512 \times 512$ image and different numbers of bins.

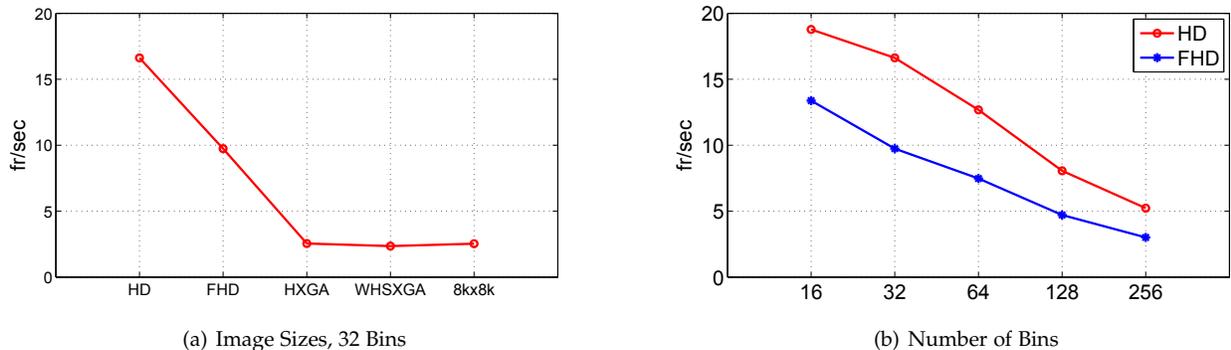

(a) Image Sizes, 32 Bins

(b) Number of Bins

Fig. 16: Frame rate when computing (a) 32-bins integral histogram for different high definition large standard images (b) integral histogram for HD and FHD images with different number of bins using four GeForce GTX 480 GPUs.

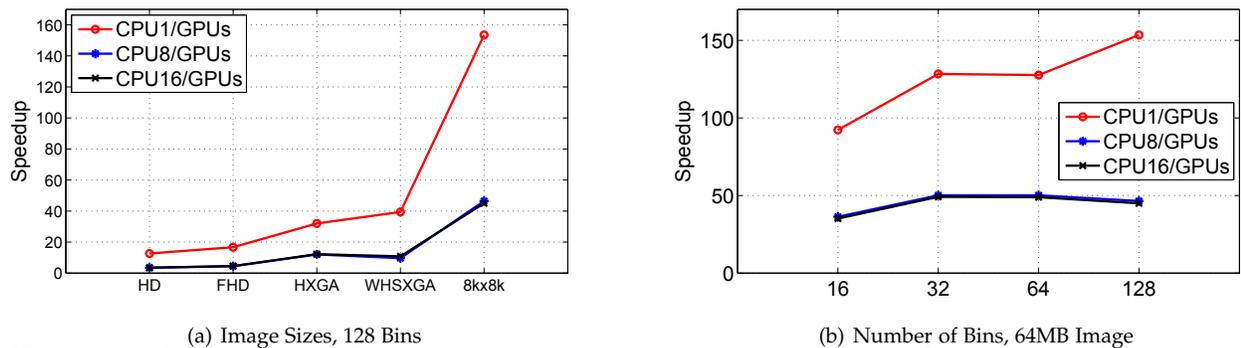

(a) Image Sizes, 128 Bins

(b) Number of Bins, 64MB Image

Fig. 17: Speedup of integral histogram computations over a CPU implementation using different degrees of multi-threading for large scale images (up to 64MB) using four GeForce GTX 480 GPUs.

- It can handle the imbalanced computation capability of heterogeneous system in which GPUs may have different hardware configuration

## 4.7   Speedup over CPU

In this section, we report the speedup of our GPU implementation of the integral histogram over a parallel CPU implementation. The speedup is defined in terms of frame rate considering data transfers between CPU and GPU. As mentioned before, these data are conservative, since in most cases the integral histogram is part of a more complex image processing pipeline, which does not require transferring the computed integral histogram back to CPU.

Figure 19 shows the speedup of GPU over CPU for image sizes varying from 256x256 to 2048x2048 and 32-bins. The GPU implementation is run on Tesla K40c GPU card. Our CPU implementation, parallelized using OpenMP, is run on an 8-core Intel Xeon E5620. Since the CPU cores are hyper-threaded, the best CPU configuration consists of 16 threads. Although the GPU implementation is data-transfer-bound for large images and/or number of bins, the speedup over the single threaded CPU implementation is about 60X, and over a 16-thread implementation, it varies between 8X and 30X for integral histogram of size $512 \times 512$ and different number of bins (Fig. 19(b)).

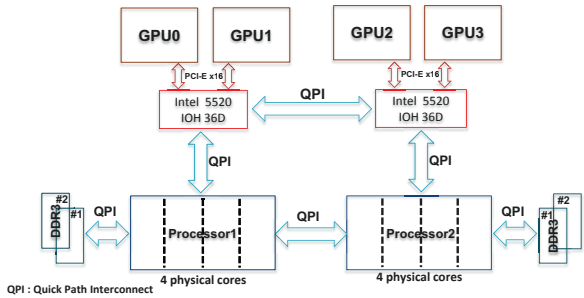

Fig. 18: General block diagram of superserver configuration

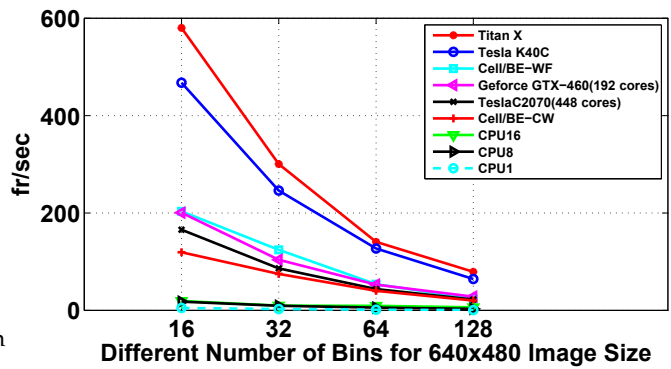

Fig. 20: Frame rate performance comparison of proposed GPU WF-TiS design versus CPU implementation using different degrees of multi-threading, CPU1, CPU8, CPU16 and Cell/B.E. performance results presented for wavefront(WF) and cross-weave(CW) scan mode using 8 SPEs in [48].

In figure 20, we have compared our WF-TiS performances utilizing several GPU cards to the best performance of the IBM Cell/B.E integral histogram parallel implementation using wave-front(WF) and cross-weave(CW) scan mode [48] for standard $640 \times 480$ image size. To calculate the frame rate of our implementation, we chose the maximum value between kernel time and data transfer time since dual buffering is applied. For our GPU implementation, in most cases the performance is data-transfer-bound: the achieved frame rate is limited by transferring the data between CPU and GPU over the PCI-express interconnect, rather than from the parallel execution on the GPU. It is shown that the *GeForce GTX Titan X* with *Maxwell* architecture outperforms all the other GPU devices, CPU and the Cell/BE processors.

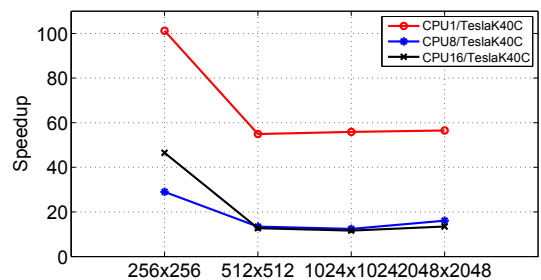

(a) Image Sizes

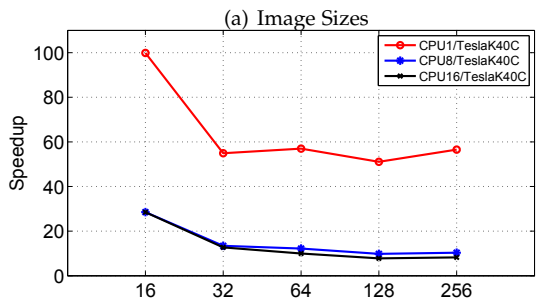

(b) $512 \times 512$ Image, Number of Bins

Fig. 19: Speedup of the GPU designs over a CPU implementation using different degrees of multi-threading

## 5 CONCLUSION

In this paper, we have evaluated four GPU implementations of the integral histogram and proposed the fastest approach to accelerate computer vision applications utilizing integral histogram. All our designs – namely CW-B, CW-STS, CW-TiS, and WF-TiS – compute parallel cumulative sums on row and column histograms either in a cross-weave or in a wavefront scan. While CW-B and CW-STS kernels are based on pre-existing scan and transpose kernels, CW-TiS and WF-TiS are based on our custom scan kernels. Our progressive optimizations were based on a careful analysis of how each alternative leverages the hardware resources offered by the GPU architecture. Our kernel optimizations coupled with the use of dual-buffering allow us to achieve a frame-rate bounded by the data transfer time over PCI-Express connecting CPU and GPU for smaller images. In particular, when computing the 32-bin integral histogram of a $512 \times 512$ image, our most efficient implementation reached a frame rate of 351 fr/sec on an GeForce GTX Titan X graphics card. However, for a sequence of larger images (HD size), which were bounded by kernel execution time, the frame rate has been improved by a factor of two for 16 bins integral histogram using dual-buffering. For large scale images, mapping integral histogram bins computations on multiple GPUs enables us to process 32 giga bytes of image data with a frame rate of 0.73 Hz. These results strengthen the idea of high performance computing to distribute the data/compute intensive tasks between multiple nodes. Furthermore, our optimized WF-TiS kernel had a 60X speedup over a serial single-threaded CPU implementation for standard image size, and a 8X to 30X speedup over a multi-threaded implementation deployed on an 8-core CPU server.